\documentclass[twocolumn,showpacs,prb,aps,superscriptaddress]{revtex4}
\usepackage{amssymb}
\usepackage{amsmath}
\usepackage{graphicx}

\begin{document}

\title{Quantum Anomalous Hall Effect in Flat Band Ferromagnet}
\author{An Zhao and Shun-Qing Shen}
\affiliation{Department of Physics and Center of Computational and Theoretical Physics,
The University of Hong Kong, Pokfulam Road, Hong Kong}

\begin{abstract}
We proposed a theory of quantum anomalous Hall effect in a flat-band
ferromagnet on a two-dimensional (2D) decorated lattice with
spin-orbit coupling. Free electrons on the lattice have
dispersionless flat bands, and the ground state is highly degenerate
when each lattice site is occupied averagely by one electron, i.e.,
the system is at half filling. The on-site Coulomb interaction can
remove the degeneracy and give rise to the ferrimagnetism, which is
the coexistence of the ferromagnetic and antiferromagnetic long-range
orders. On the other hand the spin-orbit coupling makes the band
structure topologically non-trivial, and produces the quantum spin
Hall effect with a pair of helical edge states around the system
boundary. Based on the rigorous results for the Hubbard model, we
found that the Coulomb interaction can provide an effective
staggered potential and turn the quantum spin Hall phase into a
quantum anomalous Hall phase.
\end{abstract}


\maketitle

\section{INTRODUCTION}

Quantum anomalous Hall effect (QAHE) is a quantum mechanical version
of the Hall effect in a ferromagnet in absence of an external
magnetic field or Landau levels.
Different from the quantum Hall effect in a strong magnetic field,
it originates from the
topological properties of band structure in solid.
Usually the anomalous Hall effect occurs in ferromagnetic metals due
to the intrinsic spin-orbit coupling or extrinsic spin-orbit scattering.\cite{Sinova-10RMP}
It has been realized that the Hall conductance can
be expressed as the summation of the Berry curvature in momentum space over all
occupied states of electrons.
This makes it possible to realize the
quantum Hall effect even without the Landau
levels.\cite{Haldane-88PRL} A simple picture for QAHE was proposed
in a two-dimensional electron gas with strong spin-orbit
coupling.\cite{Qi-08PRB} Electron localization in disordered systems
also provides an alternative approach to realize this
effect.\cite{Nagaosa-03PRL, Li-09PRL} The discovery of the quantum
spin Hall effect (QSHE) and topological insulators\cite{Kane-05PRL,
Kane-06PRB, Zhang-06Science, Konig-07Science, Kane-07PRL,
Zhang-09Nature, Zhang-09Science} stimulated extensive interest to
search QAHE in realistic systems. As a result of time-reversal
symmetry breaking in topological insulators, this effect was
predicted in a magnetically doped Hg$_{1-y}$Mn$_{y}$Te quantum
well\cite{Liu-08PRL} and Cr or Fe doped topological
insulators,\cite{Yu-10Science} in which the presence of
magnetization suppresses one of helical edge states in QSHE, and
preserves a chiral edge state for the quantum Hall effect. Very recently HgCr$_2$Se$_4$ 
is proposed to be a Chern semimetal and exhibits QAHE in quantum-well structure.\cite{Zhong-11PRL} 
Interaction-driven topologically non-trivial Mott insulating phase
displaying QAHE or QSHE has also attracted much research
interest.\cite{Raghu-08PRL, Ran-09PRB, Sun-09RPL}

While in search of ferromagnetism in diluted magnetic
semiconductors, rigorous models of ferromagnetism\cite{Tasaki-92PRL,
Mielke-93CMP} or ferrimagnetism \cite{Lieb-89PRL, Shen-94PRL} in
strongly correlated electron systems provides an alternative context
to realize ferromagnetism. A common feature of these examples is the
flat or almost flat band of electrons. The Coulomb interaction may
remove high degeneracy of electrons in the band, leading to the
ferromagnetism according to the Stoner criteria. Recently it was predicted
that the ferromagnetism may occur in the Hubbard model with
topological non-trivial bands.\cite{Katsura-10EPL}

In this paper we propose a theory of QAHE in a flat band ferromagnet
with spin-orbit coupling. We start with a two-dimensional decorated
lattice model, which has a pair of flat bands in the middle of the energy spectrum. The inclusion
of spin-orbit coupling makes the energy bands topologically
non-trivial, and gives rise to QSHE. Based on the rigorous
results for the Hubbard model, the Coulomb interaction may induce
the ferrimagnetism in the ground state, when the middle flat bands
are half filled. It provides an ideal way to realize a magnetic
staggered field. The staggered field can modulate the topological
numbers of electron bands by closing and reopening the energy
gap.\cite{Chu-10PRB} Different configurations of topological invariants
assigned to the electron bands can then be obtained and will give
different topological phases. Based on a self-consistent mean field
calculation, we present the phase diagram of the ground state. We
find that QAHE with nonzero Chern number can be realized in a
ferromagnet due to the Coulomb interaction.

\section{A DECORATED\ LATTICE\ MODEL}

\begin{figure}[tbp]
\includegraphics[width=0.22\textwidth]{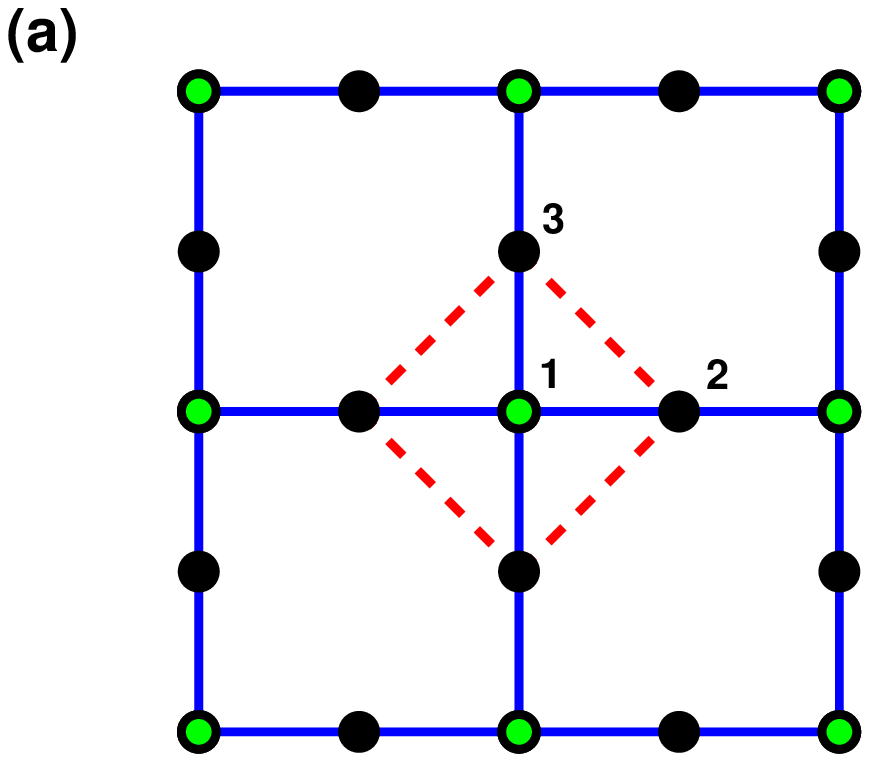} \includegraphics[width=0.22
\textwidth]{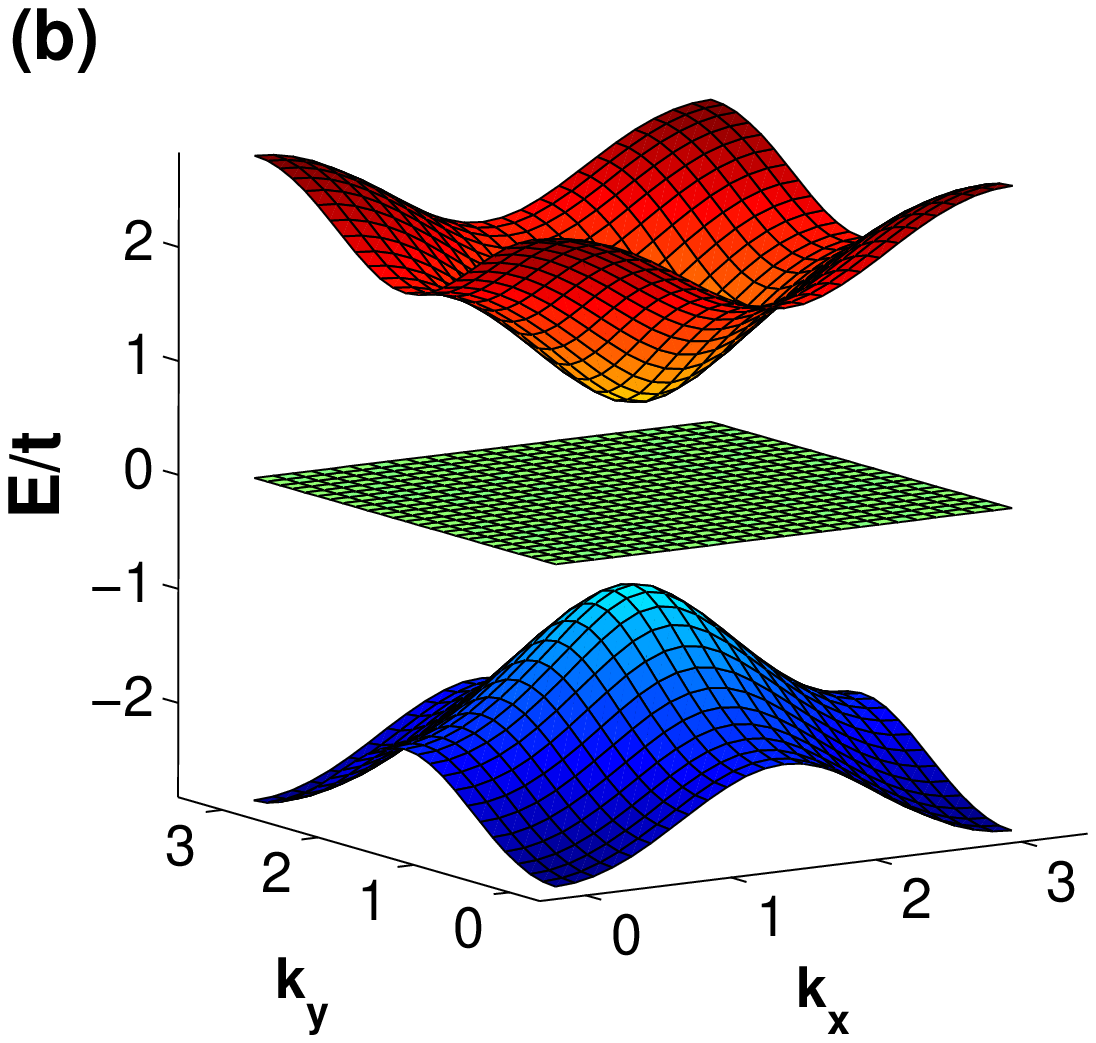} \caption{(Color online) (a) A two-dimensional decorated
lattice. Dashed lines represent the hopping with spin-orbit coupling. The light (green)
and dark (black) dots denote the sublattices $\mathbb{A}$ and $\mathbb{B}$, respectively.
(b) The energy dispersion for
$\mathcal{H}(\mathbf{k})$, which has three pairs of doubly degenerate
bands, and the gaps open in the presence of spin-orbit coupling ($\lambda \neq 0$).} \label{Fig01}
\end{figure}

We begin with the tight-binding Hamiltonian on a two-dimensional
decorated lattice [see Fig. \ref{Fig01}(a)],
\begin{equation}\label{eq000}
\mathcal{H}=\mathcal{H}_{0}+\mathcal{H}_{SO},
\end{equation}
where the spin-independent hopping term is given by
\begin{equation*}
\mathcal{H}_{0}=t\sum_{\langle i,j\rangle}c_{i}^{\dagger}c_{j},
\end{equation*}
$c_{i}=(c_{i,\uparrow},c_{i,\downarrow})^{\mathrm{T}}$ and
$c_{i,\uparrow(\downarrow)}^{\dagger}$ are the annihilation and creation
operators of electron with spin $\uparrow(\downarrow)$ on site $i$.
$\left\langle i,j\right\rangle $ means the summation over the nearest-neighbor sites. $t$ is the hopping amplitude.
The spin-orbit coupling term has the form
\begin{equation*}
\mathcal{H}_{SO}=\,\mathrm{i}\,\lambda\sum_{i\in\mathbb{A}}\sum_{j,l\in%
\mathbb{B}}c_{j}^{\dagger}\left[(\mathbf{d}_{ij}\times\mathbf{d}%
_{il})\cdot\mbox{\boldmath\ensuremath{\sigma}}\right]c_{l},
\end{equation*}
which gives a spin-dependent hopping between the next-nearest-neighbor sites with hopping amplitude $\lambda$ [shown by the dash
lines in Fig. \ref{Fig01}(a)]. Here the lattice is divided into two
sublattices, $\mathbb{A}$ and $\mathbb{B}$, shown by the light and dark dots in Fig. \ref{Fig01}(a), respectively.
$j$ and $l$ denote the adjacent sites of
site $i$, and $\mathbf{d}_{ij}$ is the unit vector along the direction from site $i$ to
site $j$, and $\mbox{\boldmath\ensuremath{\sigma}}$ are the Pauli
matrices.

We choose the sites 1, 2, and 3 in Fig. \ref{Fig01}(a) as the unit
cell. Since the $z$ component of spin $\sigma _{z}$ commutes with this
Hamiltonian, the
Hamiltonian has a block-diagonalized form in momentum space after the Fourier transformation,
\begin{equation}
\mathcal{H}=\sum_{\mathbf{k}\sigma }\Psi _{\mathbf{k}\sigma }^{\dagger
}H_{\sigma }(\mathbf{k})\Psi _{\mathbf{k}\sigma },
\end{equation}
where $\Psi _{\mathbf{k}\sigma }=(c_{1,\mathbf{k}\sigma
},c_{2,\mathbf{k} \sigma },c_{3,\mathbf{k}\sigma })^{\mathrm{T}}$,
$\sigma =\,\uparrow ,\downarrow $ denotes different spins, and
\begin{equation*}
{\small H_{\uparrow }(\mathbf{k})=\left(
\begin{array}{ccc}
0 & 2t\cos k_{x} & 2t\cos k_{y} \\
2t\cos k_{x} & 0 & -4\mathrm{i}\lambda \sin k_{x}\sin k_{y} \\
2t\cos k_{y} & 4\mathrm{i}\lambda \sin k_{x}\sin k_{y} & 0%
\end{array}
\right)}.
\end{equation*}
$H_{\downarrow }(\mathbf{k})=H_{\uparrow }^{\ast }(-\mathbf{k})$ is
the time-reversal partner of $H_{\uparrow }$. The Brillouin zone
spans over $0\leq k_{x}\leq \pi $ and $0\leq k_{y}\leq \pi $. This
Hamiltonian preserves time-reversal symmetry, i.e., $\Theta H(\mathbf{k})\Theta
^{-1}=H(-\mathbf{k})$, where $\Theta = \mathrm{i}\sigma _{y}K$ and $K$ is the complex conjugate operator.
In this spin-$1/2$ system, the time-reversal operator satisfies $\Theta ^{2}=-1$.
When $\lambda \neq 0$, in each $H_{\sigma }$, all the three bands are
well separated and can be characterized by the Chern
number\cite{Qi-08PRB}
\begin{equation}
\mathbb{C}(n,\sigma )=\frac{1}{2\pi }\int_{\text{BZ}}\mathrm{d}^{2}k\,\left[
\,\nabla \times \mathbf{\mathcal{A}}_{n}^{\sigma }(\mathbf{k})\,\right] _{z},
\end{equation}
where $\mathbf{\mathcal{A}}_{n}^{\sigma
}(\mathbf{k})=-\mathrm{i}\langle u_{n}^{\sigma
}(\mathbf{k})|\mathbf{\nabla _{k}}|u_{n}^{\sigma }(\mathbf{k}
)\rangle $ and $u_{n}^{\sigma }(\mathbf{k})$ is the Bloch
function for the $n$th band of electrons with spin $\sigma
$. Since $ H_{\uparrow }$ and $H_{\downarrow }$ are time-reversal partners,
we have $\mathbb{C}(n,\uparrow
)=-\mathbb{C}(n,\downarrow )$ for each time-reversal pairs of bands,
which are degenerate due to time-reversal symmetry. Therefore, the total Chern number is zero.
In the presence of spin-orbit coupling, the
Chern numbers of the three bands in $ H_{\uparrow }$ ($H_{\downarrow
}$) are $\{\eta ,0,-\eta \} $ ($\{-\eta ,0,\eta
\}$) from top to bottom, with $\eta =\mathrm{sgn}(\lambda )$. The nonzero
difference between $\mathbb{C}(n,\uparrow )$ and $\mathbb{C}
(n,\downarrow )$ is equivalent to a non-trivial Z$_{2}$ index,
which can also be calculated explicitly.\cite
{Fukui-05JPSJ,Fukui-07JPSJ,Moore-07PRB,Kane-06PRB} When the Fermi
level is located in the gap, the non-trivial $Z_{2}$ index for the
filled pairs of bands indicates QSHE.\cite{Franz-10PRB}

\section{EFFECT OF THE STAGGERED POTENTIAL}

For an intuitive illustration on how QAHE arises in this system,
we first introduce a spin-dependent staggered potential term
\begin{equation}
\mathcal{H}_{s}=-v_{s}\sum_{i\in \mathbb{A}}\,c_{i}^{\dagger }\sigma
_{z}c_{i}+v_{s}\sum_{i\in \mathbb{B}}\,c_{i}^{\dagger }\sigma _{z}c_{i},
\label{eq:004}
\end{equation}
and a spin-independent staggered potential term
\begin{equation}
\mathcal{H}_{c}=-v_{c}\sum_{i\in \mathbb{A}}\,c_{i}^{\dagger
}c_{i}+v_{c}\sum_{i\in \mathbb{B}}\,c_{i}^{\dagger }c_{i},  \label{eq:000}
\end{equation}
respectively, where the summations run over the sublattice sites. In
momentum space, the Hamiltonian $H(\mathbf{k})$ has a
block-diagonalized form,
\begin{equation}
H(\mathbf{k})=\left(
\begin{array}{cc}
H_{\uparrow }^{\prime }(\mathbf{k}) & 0 \\
0 & H_{\downarrow }^{\prime }(\mathbf{k})
\end{array}
\right) ,  \label{eq:003}
\end{equation}
with $H_{\uparrow }^{\prime }=H_{\uparrow }+S_{+}\Delta $ and $H_{\downarrow
}^{\prime }=H_{\downarrow }-S_{-}\Delta $, where $\Delta =\mathrm{diag}
(-1,1,1)$ and $S_{\pm }=v_{s}\pm v_{c}$. When $v_{s}\neq 0$, the
time-reversal symmetry is broken and the degeneracy of the
time-reversal pair of bands is removed. In the block-diagonalized
form, we may say that $H_{\uparrow }$ feels a staggered potential of
amplitude $S_{+}$ while $H_{\downarrow }$ feels $-S_{-}$. These two
parts of the Hamiltonian can be investigated separately.

\begin{figure}[t]
\center \includegraphics[width=0.5\textwidth]{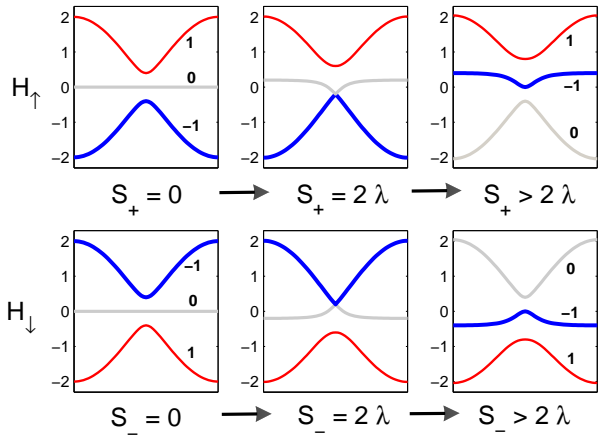} \caption{(Color
online) The band dispersions for $ H_{\uparrow
}$ and $H_{\downarrow }$ along $k_{x}\in
\lbrack 0,\protect\pi ]$, at $k_{y}=\protect\pi /2$ . The horizontal axis is $k_{x}$ and the
vertical axis is $E/t$. We take the parameter $\lambda >0$. 0 and
$\pm 1$ indicate the Chern number of each band. From the left to
right, $0<S_{\pm }<2\protect\lambda $, $S_{\pm }=2\protect\lambda $,
and $S_{\pm }>2\protect\lambda $. $H_{\downarrow }$ with $S_{-}<0$
has the identical bands structure as
$H_{\uparrow }$ with $S_{+}=-S_{-}$, but opposite Chern numbers.} \label{Fig02}
\end{figure}

We notice that a staggered potential may change the Chern numbers of
the bands of $H_{\uparrow }^{\prime }$ and $H_{\downarrow }^{\prime
}$ by closing and reopening the band gap in a band
inversion. For $H_{\uparrow }^{\prime }$, the bands cross only at
the point $\text{\textbf{k}}_{0}=(\pi /2,\pi /2)$ when $t\neq 0$,
$\lambda \neq 0$. The eigenvalues at this point are
$\{-S_{+},-4\lambda+S_{+} ,4\lambda+S_{+} \}$, respectively. As a
result, with increasing $S_+$ from zero, a band crossing happens at $
S_{+}=+2\lambda $ or $S_{+}=-2\lambda $. For example near
$S_+=2\lambda $, we may obtain an effective Hamiltonian near the point
$\text{\textbf{k}}_{0}$
\begin{equation*}
{\small \begin{aligned} H_{\textrm{eff}}=(m+\lambda q^{2})\sigma
_{z}-\sqrt{2}t(q_{x}\sigma _{x}-q_{y}\sigma _{y})+\lambda (q^{2}-2),
\end{aligned}}
\end{equation*}
where $\mathbf{q}=\mathbf{k}-\mathbf{k}_{0}$ and $m=S_{+}-2\lambda $. From
this two-band massive Dirac model it is known that the topological quantum
phase transition occurs when the sign of $m$ or $\lambda $ changes. The
Chern number of the lower band is given by\cite{Lu-10PRB} (the upper one has a sign change)
\begin{equation*}
C_{L}=-\frac{1}{2}\left[ \mathrm{sgn}(m)+\mathrm{sgn}(\lambda )\right] .
\end{equation*}
Thus the sign change of $m$ indicates that the Chern number changes
from 0 to 1 or 1 to 0. Fig. \ref{Fig02} depicts the band structure
for $H_{\uparrow }^{\prime }$ and
$H_{\downarrow }^{\prime }$, with the Chern numbers also denoted for each band.
There are three different cases: (a)
when $0<S_{+}<2\lambda $, the Chern numbers of the bands in
$H_{\uparrow }^{\prime }$ are $\{1,0,-1\}$ from top to bottom; (b)
when $S_{+}=2\lambda $, the two lower bands touch at \textbf{k}$
_{0}=(\pi /2,\pi /2)$ and the Chern numbers disappear as the two
bands are not well separated; (c) when $S_{+}>2\lambda $ the band
gap reopens and the Chern numbers become $\{1,-1,0\}$ after the two inverted bands exchange
their Chern numbers. Similarly, for
$H_{\downarrow }^{\prime }$, the Chern numbers of the three bands
are $\{-1,0,1\}$ when $0<S_{-}<2\lambda $ and $ \{0,-1,1\}$ when
$S_{-}>2\lambda $. It is noted that the band structure and Chern
numbers of $H_{\downarrow }^{\prime }$ with $S_{-}$ and $\lambda$
are identical to those of $H_{\uparrow }^{\prime }$ with $S_{+}=-S_-$
and $ -\lambda $.

\begin{figure}[b]
\includegraphics[width=0.5\textwidth]{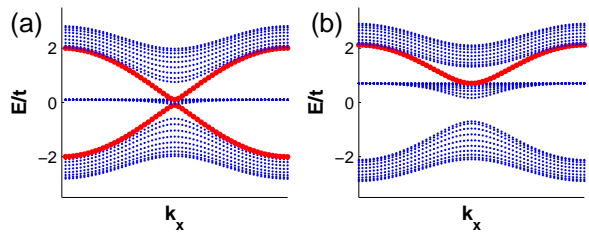}
\caption{(Color online) Band structure of $H_{\uparrow }$ in a strip geometry.
Edge state spectra are shown by dark gray lines
(or red online). (a)
$0<S_{+}<2\protect \lambda $. (b) $S_{+}>2\protect\lambda $.}
\label{Fig03}
\end{figure}

\begin{figure}[t]
\includegraphics[width=0.5\textwidth]{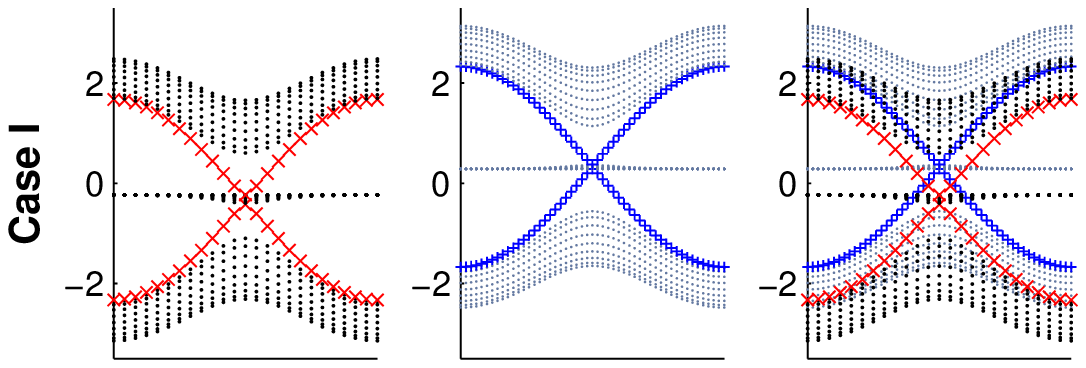}
\includegraphics[width=0.5
\textwidth]{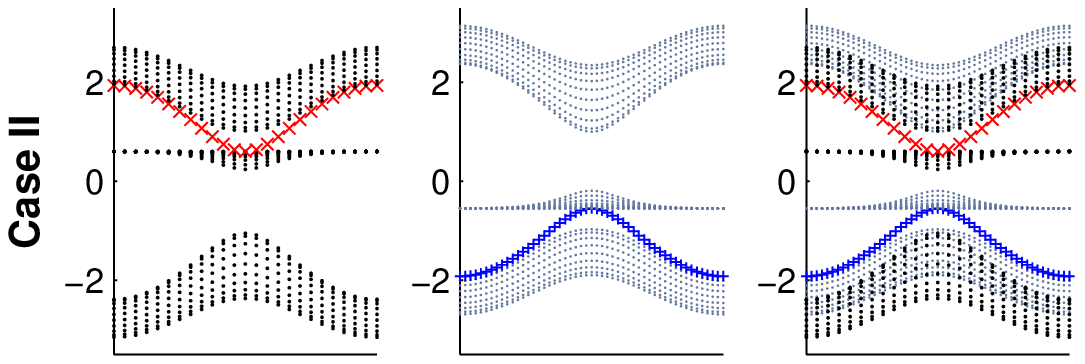} \includegraphics[width=0.5\textwidth]{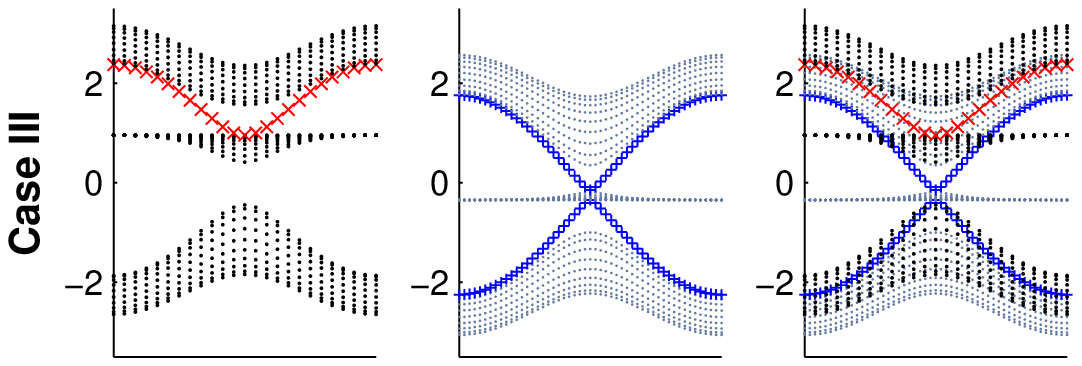}
\includegraphics[width=0.5\textwidth]{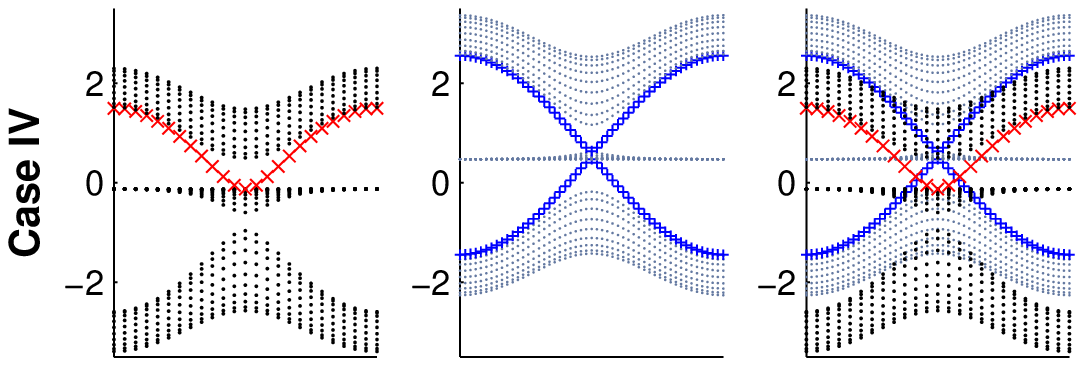}
\includegraphics[width=0.5
\textwidth]{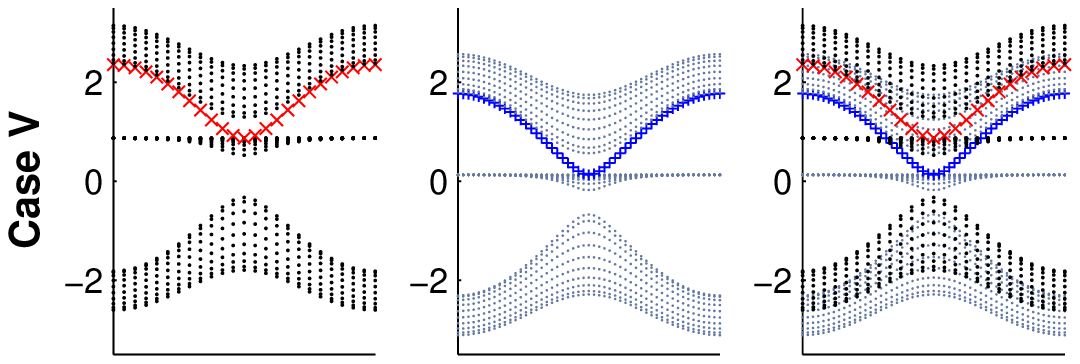} \caption{(Color online) Band structure in a strip
geometry for five different cases. Gray large crosses and dark gray
small crosses (red and blue online) represent the dispersions for the edge
states. The columns from left to right are for
$H_{\uparrow }^{\prime }$, $H_{\downarrow }^{\prime }$ and $H_{\uparrow }^{\prime} + H_{\downarrow
}^{\prime }$, respectively. The horizontal axis is $k_{x}\in[
0,\protect\pi]$ and the vertical axis is $E/t$.} \label{Fig04}
\end{figure}

According to the Chern numbers, the edge-bulk correspondence tells
that the edge states in a sample of strip geometry depend on parameters.
In Fig. \ref{Fig03}, we only present
the edge state spectra of $H_{\uparrow }^{\prime }$. When
$0<S_{+}<2\lambda$, the edge state spectra connect the upper and
lower bands. Since the Chern number of the
middle band is zero, this topologically trivial band only distorts
the edge state spectra, but does not affect the existence of the edge
states. The total Chern number is still 1
and the system is topologically non-trivial when the middle and
lower bands are fully filled. When $S_{+}>2\lambda $, the middle
band becomes topologically non-trivial, and the lower band becomes
trivial. The edge state spectra only connect the middle and the
upper bands.

Now we can present the evolution of edge states in the total Hamiltonian
$H_{\uparrow }^{\prime} + H_{\downarrow
}^{\prime }$. Five cases are listed in Fig. \ref{Fig04}. Without loss of
generality, we take $v_{s}>0$ and $v_{c}\geq 0$. In this case we have $
S_{+}\geq |S_{-}|$. When time-reversal symmetry is broken, an
energy gap can open between a time-reversal pair of bands. A uniform
magnetism term $M\sigma _{z}$ is introduced to shift the bands of $H_{\uparrow
}^{\prime }$ and $H_{\downarrow }^{\prime }$ upward and downward, respectively,
without changing the Chern number of each band. A
gap is opened between two middle bands. At half filling,
we assume that the Fermi level is located in this gap. According to the
Chern numbers and relative positions of the energy bands, the system
can be categorized into five cases. Case I: $S_{+}<2|\lambda |$
and $|S_{-}|<2|\lambda |$. The total Chern numbers of three lower
bands is $0 $. However, the nonzero difference between the Chern numbers of
$H_{\uparrow }^{\prime }$ and $H_{\downarrow }^{\prime}$ indicates QSHE.
We may have two counter-propagating edge
states with different spins on each edge,
although these two edge states do not form a time-reversal pair as
time-reversal symmetry has already been broken. Case II:
$S_{+}>2|\lambda |$ and $S_{-}>2|\lambda |$. The Chern numbers of
both $H_{\uparrow }^{\prime }$ and $H_{\downarrow }^{\prime
}$ change to zero due to the $v_{s}$ term. When the middle band of $
H_{\uparrow }^{\prime }$ is higher than the middle band of $H_{\downarrow
}^{\prime }$, the system is in an insulating phase as shown in the figure.
When the middle band of $H_{\uparrow }^{\prime }$ is lower, the system exhibits
QSHE. Case III and Case IV: $S_{+}>2|\lambda |$ and $|S_{-}|<2|\lambda |$.
Case III, when the middle band of $H_{\uparrow }^{\prime }$ is higher than the
middle band of $H_{\downarrow }^{\prime }$, the total Chern number is $1$.
This is a QAHE phase, in which there exists one gapless spin-up chiral edge state.
Case IV: when the middle band of $H_{\uparrow }^{\prime }$ is
lower than the middle band of $H_{\downarrow }^{\prime }$, the total Chern
number is $0$, but it gives rise to QSHE. To distinguish Case III and Case
IV, one can check the eigenvalues at $(\pi /2,\pi /2)$, and we have Case III
when $v_{c}+M>2|\lambda|$. Case V: $S_{+}>2|\lambda |$ and $
S_{-}<-2|\lambda |$. Due to $v_{c}$ term, the Chern numbers of both $H_{\uparrow }^{\prime }$ and
$H_{\downarrow }^{\prime }$ change, and the total Chern
number is $1$. Once again the system exhibits QAHE as there is only one
chiral edge state.

\section{FERRIMAGNETSIM AND QAHE}

One of the prominent features of the model in Eq.\eqref{eq000} is the
appearance of the flat bands due to the unequal numbers of sites of the two
sublattices, even in the presence of spin-orbit coupling.
These flat bands give rise to the
famous flat-band ferromagnetism when the Coulomb interaction is turned on.
\cite{Tasaki-92PRL, Mielke-93CMP} When the system is at half filling,
the two lower bands are fully filled, and the total spin of these
two bands is zero since electrons in these two bands have opposite
spins. The two middle bands are degenerate. In this case, if only
one single middle band is fully filled, the expectation value of the
Coulomb interaction is minimized since the fully polarized electron spin in the middle band excludes the double occupancy
completely at each site. In this way, the ground state of the system
is ferromagnetic. The total spin is given by the degeneracy of the
flat band, $S_{tot}=N\hbar /2$.\cite{Lieb-89PRL, Shen-98IJMP}
Furthermore, since the antiferromagnetic correlation is
dominant in the half-filled Hubbard model,
this ground state is actually ferrimagnetic in which ferromagnetic
and antiferromagnetic long range orders coexist.\cite{Shen-94PRL}
When the spin-orbit coupling is present, the flat bands will be
distorted by the ferrimagnetism. When the coupling is strong, the
ferrimagnetism would significantly distort the flat bands and is suppressed.
Let us focus on the case of weak spin-orbit
coupling, in which the band is expected to be almost flat. It is
still possible that the ferrimagnetism could survive if the Coulomb
interaction is strong enough over the band
distortion.\cite{Tasaki-96JSP} Thus the combination of the flat band
and the Coulomb interaction provides a reliable mechanism to realize
the spin-dependent staggered potential in this system.

The Hamiltonian with the on-site and nearest-neighbor repulsive interactions
has the form
\begin{equation}
\begin{aligned}\mathcal{H}=\; & \mathcal{H}_0 + \mathcal{H}_{\textrm{SO}} +
U\sum_{i}n_{i\uparrow}n_{i\downarrow}+ \\ &V\sum_{\left\langle
i,j\right\rangle
}(n_{i}-1)(n_{j}-1)-\mu\,(\,\sum_{i}n_{i}-3N\,),\end{aligned}
\end{equation}
where $U>0$ is the on-site Coulomb potential, $V>0$ is the nearest-neighbor
repulsive potential, $n_{i\alpha }=c_{i\alpha }^{\dagger }c_{i\alpha }$ is
the number operator for electron with spin $\alpha $ on site $i$,
and $n_{i}=n_{i\uparrow }+n_{i\downarrow }$. The chemical potential $\mu $
determines the number of electrons in the system. $N$ here is the number of
unit cells, and the number of electrons is $3N$ at
half filling. In the mean field approximation, the on-site interaction is
decoupled as\cite{Note-MF}
\begin{equation}
Un_{i\uparrow }n_{i\downarrow }\approx -Um_{i}\left( c_{i}^{\dagger }\sigma
_{z}c_{i}\right) +\frac{U}{2}m_{i}^{2}+\frac{U}{2}n_{i},
\end{equation}
where $m_{i}\equiv \langle c_{i}^{\dagger }\sigma _{z}c_{i}\rangle $ is the
magnetization on site $i$. Due to the asymmetry of the two sublattices, the
magnetization on site A and B are different, saying, $m_{A}$ for the sublattice
$\mathbb{A}$ and $m_{B}$ for the sublattice $\mathbb{B}$. In this way the
Hubbard term is reduced to
\begin{equation}
\sum\limits_{i}Un_{i\uparrow }n_{i\downarrow }=-Um_{A}\sum_{i\in \mathbb{A}
}\,c_{i}^{\dagger }\sigma _{z}c_{i}-Um_{B}\sum_{i\in \mathbb{B}
}\,c_{i}^{\dagger }\sigma _{z}c_{i},
\end{equation}
At half filling, since the antiferromagnetic correlation is
dominant, $m_{A}$ and $m_{B}$ have opposite signs, based on the
rigorous results for the Hubbard model.\cite{Shen-94PRL,Shen-98IJMP}
This can also be illustrated from the calculation of the mean field
theory.

The nearest-neighbor interaction may induce the instability of
charge-density wave (CDW). In the mean field approximation, we have
\begin{eqnarray}
&&V\sum_{\langle i,j\rangle}(n_{i}-1)(n_{j}-1)\nonumber\\
&\approx & 4V\sum_{i\in A}\rho_{B}(n_{i}-1)+2V\sum_{i \in B}\rho_{A}(n_{i}-1)-4VN\rho_{A}\rho_{B},\nonumber\\
\end{eqnarray}
where $\rho _{A}\equiv \langle n_{i}-1\rangle $ for the sublattice $\mathbb{A
}$ and $\rho _{B}\equiv \langle n_{i}-1\rangle $ for the sublattice $\mathbb{
B}$. The CDW order parameter $\rho $ is given by $\rho_{A}=2\rho $
and $\rho_{B}=-\rho $ due to the charge conservation.

\begin{figure}[t]
\center \includegraphics[width=0.45\textwidth]{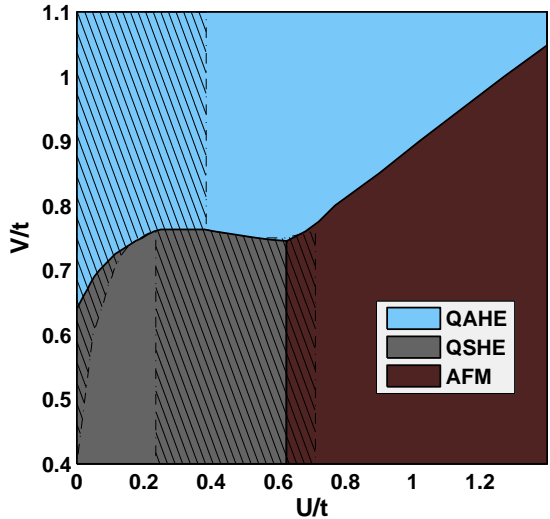}
\caption{(Color online) $U-V$ phase diagram for $\protect\lambda
=0.1t$ at half filling. The shadow marks the gapless regions at
half filling, which should be metallic with strong anomalous Hall
effect and spin Hall effect.} \label{Fig05}
\end{figure}
After some tedious algebra we may have the zero-temperature mean field free energy at half filling
\begin{equation}
{\small \begin{aligned} & F(m_{A},m_{B},\rho,\mu)\\ = &
\;\sum_{i,\mathbf{k}}\theta\left[\mu-E_{i}(\mathbf{k})\right]\left[-
\mu+E_{i}(\mathbf{k})\right]+N(E_{0}+3\mu),\end{aligned}}
\end{equation}
where $\theta (x)$ is the step function, $E_{i}(\mathbf{k})$ is the
eigenvalues of $H(\mathbf{k})$ in Eq. \eqref{eq:003} with $H_{\uparrow
}^{\prime }=H_{\uparrow }+(v_{s}+v_{c})\Delta +M$ and $H_{\downarrow
}^{\prime }=H_{\downarrow }-(v_{s}-v_{c})\Delta -M$, where $v_{s}=U(m_{A}-m_{B})/2$, $v_{c}=4\rho V$, and $M=-U(m_{A}+m_{B})/2$.
$E_{0}=Um_{A}^{2}/2+Um_{B}^{2}+8V\rho ^{2}-4V\rho $. The summation runs over
the whole Brillouin zone. The order parameters $m_{A}$, $m_{B}$,
$\rho$, and $\mu $ can be determined self-consistently by minimizing the free energy. The
variational principle
\begin{equation}
\frac{\delta F}{\delta m_{A}}=\frac{\delta F}{\delta m_{B}}=\frac{\delta F}{
\delta \rho }=\frac{\delta F}{\delta \mu }=0
\end{equation}
leads to a set of the mean field equation,
\begin{equation*}
{\small \begin{aligned} &\sum_{i,\mathbf{\mathbf{k}}}\frac{\partial
E_{i}(k)}{\partial m_{A}}\theta (\mu -E_{i}(\mathbf{k}))+NUm_{A}=0, \\
&\sum_{i,\mathbf{k}}\frac{\partial E_{i}(\mathbf{k})}{\partial m_{B}}\theta
(\mu -E_{i}(\mathbf{k}))+2NUm_{B}=0, \\ &\sum_{i,\mathbf{k}}\frac{\partial
E_{i}(\mathbf{k})}{\partial \rho }\theta (\mu -E_{i}(\mathbf{k}))+NV(16\rho
-4)=0, \\ &\sum_{i,\mathbf{k}}\theta (\mu -E_{i}(\mathbf{k}))-3N=0. \\
\end{aligned}}
\end{equation*}
We solve this set of equations numerically. The calculated mean field results are consistent with the rigorous results for the Hubbard model. $m_A$ and $m_B$ have different signs, which demonstrates the existence of antiferromagnetic correlation. $m_A+2m_B\neq 0$ demonstrates the ferromagnetic correlation. If $v_c=0$, $m_A+2m_B=1$, which is one of the rigorous results for the Hubbard model.

Figure. \ref{Fig05} shows the mean field $U-V$ phase diagram for
$\lambda =0.1t$ and $t>0$. The CDW order is zero when $V$ is small.
$v_{s}$ and $M$ would increase with $U$ and have the same sign.
The system transits from Case I to Case II through a band inversion.
However we may see that the gap at half filling is not opened in
the dashed area of Fig. \ref{Fig05}. When this gap opens, Case I gives
an AFM quantum spin hall effect, and Case II is an AFM insulating
phase. When $V$ is large, the CDW order may become nonzero and increase
with $V$. Case III can be found near the transition point
where $v_{c}$ is small, and Case V can be found at larger $V$ where
$v_{c}$ becomes large. They both present QAHE when the gap opens
at half filling. If large $\lambda$ is chosen, the $v_{c}$ term would
distort flat bands so much that the orders of $v_{s}$ and $M$ are
suppressed. Thus to have a QAHE, we need strong on-site interaction
$U$ and a nonzero spin-independent staggered field.

\section{CONCLUSIONS}

We have found that the flat-band ferromagnet may exhibit QAHE after the
inclusion of spin-orbit coupling on a 2D decorated lattice. The
spin-orbit coupling can induce topologically non-trivial phase on
this lattice, which exhibits QSHE. In the present three-band system,
the existence of the topologically trivial flat band between the two
non-trivial bands does not affect the formation of QSHE. The Coulomb
interaction may remove the degeneracy of electrons in the flat band
and lead to spontaneous symmetry breaking, which gives rise to
ferrimagnetism. The coexistence of ferrimagnetism and CDW may break
the balance between the helical edge states with spin up and spin
down in QSHE, and make it possible that one branch of edge states
is suppressed completely, and the other survives. As a result,
it gives rise to QAHE.

\acknowledgments This work was supported by the Research Grant Council of
Hong Kong under Grant Nos. HKU 7037/08P and HKUST3/CRF/09.


\begin{thebibliography}{99}
\bibitem{Note-MF} In the mean field approximation,
\begin{eqnarray*}
Un_{i\uparrow }n_{i\downarrow } &=&U(\left\langle n_{i\uparrow
}\right\rangle +\delta n_{i\uparrow })(\left\langle n_{i\downarrow
}\right\rangle +\delta n_{i\downarrow }) \\
&\approx &U\left(\left\langle n_{i\uparrow }\right\rangle \left\langle
n_{i\downarrow }\right\rangle +\delta n_{i\downarrow }\left\langle
n_{i\uparrow }\right\rangle +\delta n_{i\uparrow }\left\langle
n_{i\downarrow}\right\rangle\right)
\end{eqnarray*}%
and the term $U\delta n_{i\uparrow }\delta n_{i\downarrow }$ is droped as a
higher order term. Define $m=\left\langle n_{i\uparrow }\right\rangle
-\left\langle n_{i\downarrow }\right\rangle $. Then $\left\langle
n_{i\uparrow }\right\rangle =\frac{1}{2}(1+m)$ and $\left\langle
n_{i\downarrow }\right\rangle =\frac{1}{2}(1-m)$ when the system is half
filled.
\end{thebibliography}
\end{document}